\documentclass{ws-mpla}
\usepackage[super]{cite}
\usepackage{graphicx}
\usepackage{amsfonts}
\usepackage{upgreek}
\begin{document}

\markboth{G. Mandaglio et al.}
{First measurement of the 2.4 and 2.9 MeV $^6$He three-cluster resonant states...}

\catchline{}{}{}{}{}

\title{First measurement of the 2.4 and 2.9 MeV $^6$He three-cluster resonant states via the $^3$H($^4$He,p$\alpha$)2n four-body reaction}

\author{\footnotesize Giuseppe Mandaglio$^{1,2,3,*}$, Orest Povoroznyk$^{4,\dagger}$, Olga K. Gorpinich$^{4}$,
  Olexiy O.Jachmenjov$^{4}$, Antonio Anastasi$^{2,3}$, Francesca Curciarello$^{2,3}$, Veronica De Leo$^{2,3}$,   Hanna V. Mokhnach$^{4}$, Oleg Ponkratenko$^{4,\dagger}$, Yuri  Roznyuk$^{4}$, Giovanni Fazio$^{2,3}$, and Giorgio Giardina$^{2,3,\ddagger}$ }

\address{$^1$Centro Siciliano di Fisica Nucleare e Struttura della Materia, 95125 Catania, Italy \\
$^2$Dipartimento di Fisica e di Scienze della Terra dell'Universit\`a di Messina, 98166 Messina, Italy \\
$^3$Istituto Nazionale di Fisica Nucleare, Sezione di Catania, 95123 Catania, Italy\\
$^4$Institute for Nuclear Research, National Academy of Science of Ukraine, 03680 Kiev, Ukraine\\
$^*$gmandaglio@unime.it,$^\dagger$orestpov@kinr.kiev.ua,$^\ddagger$ggiardina@unime.it}

\maketitle

\pub{Received (Day Month Year)}{Revised (Day Month Year)}

\begin{abstract}
Two new low-lying $^6$He levels at excitation energies of about 2.4 and 2.9 MeV were observed in the experimental investigation of the p-$\alpha$ coincidence spectra obtained by the $^3$H($^4$He,p$\alpha$)2n four-body reaction at $E_{\rm \,^4He}$ beam energy of 27.2 MeV. The relevant $E^*$ peak energy and  $\Gamma$ energy width  spectroscopic parameters for such $^6$He$^*$ excited states decaying into the $\alpha$+n+n channel were obtained by analyzing the bidimensional ($E_{\rm p}$,~$E_{\rm \alpha}$) energy spectra. The present new result of two  low-lying $^6$He$^*$ excited states  above the $^4$He+2n threshold energy of  0.974 MeV  is important for  the investigation of the nuclear structure of neutron rich light nuclei and also as a basic test for theoretical models in the study of the three-cluster resonance feature of $^6$He.

\keywords{low-lying $^6$He levels, tritium target, four-body reaction}
\end{abstract}

\ccode{PACS Nos.: 27.20.+n, 25.55.-e, 24.30.-v}

\section{Introduction and motivation}	

\begin{figure}
\vspace{0.5cm}
\centering{\resizebox{0.4\textwidth}{!}{%
  \includegraphics{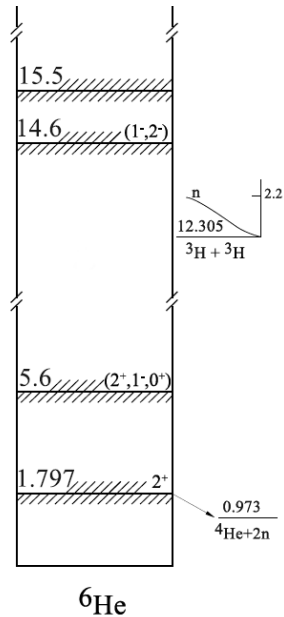}}\vspace{-0.3cm}
}
\caption{Energy level diagram of  $^6$He in the Tilley {\it et al.}\cite{till02} compilation.}
\label{f1levels}       
\end{figure}

Among the lightest nuclei, research of the structure and decay of the $^6$He states is one of the most intriguing phenomena of modern nuclear physics.  Numerous theoretical studies (see the Ajzenberg-Selove \cite{selov88} and Tilley {\it et al.} \cite{till02}  compilations) were directed towards determining the resonant structure nature of  $^6$He. The low-lying $^6$He$^*$ excited states decay into the $^4$He+2n three-cluster channel, whose threshold energy is 0.974 MeV,  while the high-lying $^6$He$^*$ excited states appear to decay into the $^3$H+$^3$H two-cluster channel, whose threshold energy is 12.305 MeV. Moreover, in the interval between these two threshold energies for the $\alpha$+2n and t+t decay channels,  the compilation \cite{selov88} shows only one narrow 1.8 MeV excited state while the compilation  \cite{till02} also shows the presence of the 5.6 MeV excited state (see Fig. \ref{f1levels}) with a very large $\Gamma$ width of 12.1 MeV. At first, these low-lying excited levels were examined as soft dipole resonances \cite{Hansen,Suzuki} consisting of an oscillation of two neutrons with respect to the $\alpha$-cluster. In subsequent theoretical calculations  \cite{Csoto,Danilin,Thompson} the above-mentioned $^6$He$^*$ excited  states were considered to be three-body continuum states. In addition, 
many experimental and theoretical studies\cite{Navratil,Pieper,Volya,Hagen,Myo} were devoted to study the energy levels in $^6$He, but the obtained results are quite controversial for the excited states above the 2$^+$  level at 1.797 MeV.

 The experimental investigations of the low-lying excited states of $^6$He   were concentrated on the analysis of inclusive spectra measured in charge-exchange  $^6$Li($^7$Li,$^7$Be)$^6$He  \cite{Hansen,Janecke,Annakkage,Nakayama00,Sakuta} and $^6$Li(t,$^3$He)$^6$He \cite{Nakamura00} reactions, while in the recent p($^8$He,t) reaction\cite{mougeot12} the low-lying $^6$He$^*$ excited states were investigated by an inverse kinematic experiment via a 2-neutron transfer reaction, by analyzing the energy spectra of tritons. 
 
In the experiments where the information on the excitation of missing nuclei is extracted from inclusive spectra a correct procedure of accounting for the non-resonant background and all resonance contributions due to other reaction channels is needed. At the same time, the  analysis carried out in papers \cite{Janecke,Nakamura00} contained the series of contradictory suppositions in relation to the selection and accounting for the background contributions. For example, the non-resonance background in the ($^7$Li,$^7$Be)  reaction was calculated but it was not measured, while in paper   \cite{Nakayama00} the  background contributions were not taken into account in general. In order to remove the background contributions it is necessary to consider the degrees of freedom connected with the excitation of both nuclei in the incident channel and  reaction products in the exit channel. 
The $^6$He  nucleus has a distinctive three-cluster feature at low-energy excited states. It has one loosely bound ground state with the energy of -0.974 MeV with respect to the $\alpha$+n+n three-cluster threshold and there are no bound states in any two-cluster subsystems. 

Another light nucleus with a strong three-cluster features is $^6$Be, and  the investigation of  the three-cluster resonance nature of the $^6$He and $^6$Be nuclei is particularly important because it is possible to obtain information on the role of the large excess of neutrons in the case of the $^6$He nucleus and the large excess of protons in the case of $^6$Be, in connection with the characteristics and nature of the low-lying three-cluster resonance states for these two  mirror nuclei.  On the other hand, it is also interesting to promote  experimental investigations of the low-lying excited state of $^6$Be (mirror nucleus of $^6$He) where no threshold energy exists for the $\alpha$+p+p three-cluster formation.

An  appropriate and elaborate procedure was used in the investigation of  the photonuclear $^7$Li($\gamma$, p)$^6$He reaction \cite{Boland} by using the population of excited levels created in the photonuclear reaction\cite{owens}.
In addition,    compilation \cite{till02} gives  very broad $\Gamma$ width values for $^6$He$^*$ excited states up to high $E^*$ excitation energies (see for example the papers\cite{Akimune03,Yamagata}), only with the exclusion of the very narrow low-lying first excited state at $E^*=1.797$~MeV. Therefore,
 we decided to investigate the low-lying excited $^6$He$^*$ states just above the  0.974 MeV threshold energy value populating the $\alpha$+2n channel in order to observe and resolve the presence of other possible low-lying  $^6$He$^*$ excited states by analysis of the ($E_{\rm p}$, $E_{\rm \alpha}$) bidimensional coincidence spectra of the $^3$H($^4$He,p\,$\alpha$)2n reaction.  To this aim  we needed to work with a relatively low beam energy, but high enough to explore the excited $^6$He$^*$ states just above the mentioned $\alpha$+2n threshold energy value, because by increasing the beam  energy  the contributions from other concurrent channels also increase. Therefore, higher beam energies lead to a worse determination of the $E^*$ and $\Gamma$ widths. In such cases, it is impossible to resolve two neighboring  distinguished $^6$He$^*$ excited states. In fact, some difficulties also exist in the analysis of two particle coincidence spectra if   all resonant contributions due to the concurrent reaction channels that lead to the same final products are not considered and resolved.

Moreover, in paper \cite{pov12} we discussed  the difficulty of reaching reliable $E^*$ and $\Gamma$ measurements by analysis of single (inclusive) particle spectra due to the huge and difficult estimation of the complete background present in such spectra.  In fact, in the analysis of single particle spectra,  the modest yield of a possible resonant contribution may not appear, and one might observe   only a  large single resonant yield instead of a resolved feature without an adequate analysis of the various contributions due to the concurrent channels. In the latter case, the single large resonant contribution appears as the convolution of various resonant contributions due to the population of : i) different excited states of the same nucleus,  ii) excited states of other nuclei formed in intermediate steps of the reaction, and iii)   other reaction mechanisms, but all leading  to the same final products.

 By taking into account all of the above-mentioned reasons, we decided to study the low-lying $^6$He$^*$ excited states by analyzing the ($E_{\rm p}$, $E_{\rm \alpha}$) bidimensional coincidence spectra of the   $^3$H($^4$He,p\,$\alpha$)2n four-body reaction at $E_{\rm ^4He}$ beam energy of 27.2 MeV produced by the Y-120 cyclotron accelerator. At this beam energy we can explore the low excitation energy region of $^6$He$^*$ just above the $\alpha$+2n threshold energy with well controlled experimental and analysis conditions.

\section{Experimental Setup and Event Selection}

In order to study the $^3$H($^4$He,p$\alpha$)2n reaction 
 by the analysis of  ($E_{\rm p}$,~$E_{\rm \alpha}$) bidimensional spectra, we used the apparatus described in our previous work \cite{pov12,pov11} where the target made of  titanium backing (2.6 mg/cm$^2$ thick) saturated with tritium (equivalent to the thickness of about 0.15 mg/cm$^2$) was used, while in the present experiment the $^4$He-particle beam of $27.2 \pm 0.15$ MeV was produced by the  cyclotron accelerator Y-120 of the Institute for Nuclear Research in Kiev.  The gas of tritium diffused through the titanim backing was pure at 99.9~\%.
 The impurities of 0.1 \% can be due to the presence of hydrogen H and deuterium $^2$H. The  
 H($^4$He,p)$\alpha$ reaction is an elastic scattering process  ($Q$-value = 0~MeV), and the  $^2$H($^4$He,p$\alpha$)n reaction is a three-body (p, $\alpha$, n) reaction with $Q$-value = $-2.22$~MeV. This value is  different with respect to the studied $^3$H($^4$He,p$\alpha$)2n four-body (p, $\alpha$, 2n) reaction with $Q$-value = $-8.48$~MeV.

To detect the products of the $^4$He+$^3$H reaction and to avoid the coincidence events related to the particles present in the above-mentioned reaction that are not of our interest, we used two $\Delta E-E$ telescopes placed at the left and right sides with respect to the beam direction defined as the polar axis. Therefore, the identification and energy determination of outgoing p and $\alpha$ charged particles with an energy resolution of about 100 keV allowed us to obtain the ($E_{\rm p}$,~$E_{\rm \alpha}$) bidimensional spectra of the coincidence events. The telescope placed on the left side consisted of $\Delta E$ (25 $\mu$m thick totally depleted silicon surface barrier detector (SSD)) and E [Si(Li) with 1\,mm$^{t}$, 261 mm from the target] detectors to identify and separate $\alpha$-particles of energy greater than 4.4 MeV, while the telescope
 placed on the right side consisted of $\Delta E$ [100 $\mu$m  SSD]  and $E$ [Si(Li) with 1.5 mm$^{t}$,  250 mm  from the target] detectors to identify and separate protons of energy greater than 3.2 MeV from deuteron  and triton  particles. 
 For the energy calibration of silicon detectors   
 a standard technique was used for the SSD. The solid angles of left and right telescopes were 1.30 and 1.44 msr, respectively, with an angle resolution of about 1$^\circ$. We  recorded the signals coming from the two telescopes within a window  of about 100 ns by using a standard electronic set-up,  choosing windows on the corresponding bit-pattern and the relevant time-to-amplitude spectra. We checked that the background of the p-$\alpha$ coincidence events were completely absent when only the titanium backing   was  used.

\begin{figure}
\centering{\resizebox{0.78\textwidth}{!}{%
  \includegraphics{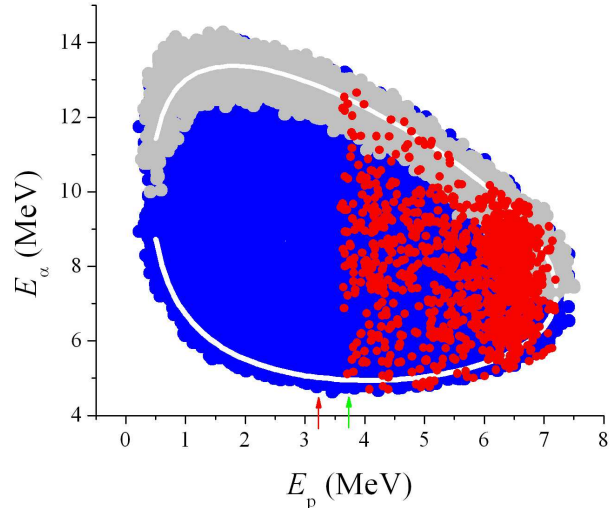}}\vspace{-0.3cm}
}
\caption{Experimental bidimensional spectrum of p-$\alpha$  coincidence events at $\theta_{\rm p}=28.5^\circ$ and $\theta_{\rm \alpha}=16.5^\circ$,     and $E_{\rm \,^4He}$ beam energy of 27.2 MeV. 
Red dots represent  the  experimental result; the solid white lines  represent  kinematic loci  of the $^3$H($^4$He,p$\alpha$)$<2n>$  three-body reaction, in the frame of a punctual geometry of detectors. Light grey
dots represent the  simulation of the upper branch of kinematic loci for the p+$\alpha+<$2n$>$ reaction  products. Blue circles represent the simulation of background due to process (\ref{eq8}). The arrow at $E_{\rm p}=3.2$~MeV indicates the lower energy limit for the proton detection.}
\label{f1}       
\end{figure}

Starting from the $^4$He+$^3$H collision in the entrance channel,  the p+$\alpha$+n+n four-body products in the exit channel can be produced via:

  \begin{eqnarray}
^4{\rm He}  + ^3{\rm H} & \rightarrow & p + ^6{\rm He}^* \rightarrow  p + \alpha + n+ n  \label{eq1} \\
              & \rightarrow & \alpha + ^3{\rm H}^*     \rightarrow   \alpha+ p + n + n  \label{eq2} \\
              & \rightarrow & n + ^6{\rm Li}^* \rightarrow  n + p + \alpha + n  \label{eq3} \\          
             & \rightarrow     &     ^5{\rm Li}^* +n +n  \rightarrow  p + \alpha + n + n  \label{eq4} \\
              & \rightarrow     &     ^5{\rm He}^* +p +n  \rightarrow  n + \alpha + p + n  \label{eq6} \\
             & \rightarrow &  p + \alpha + n + n , \label{eq8}              
\end{eqnarray}
where processes (\ref{eq1}), (\ref{eq2}), and (\ref{eq3})  are the mechanisms in which unbound resonance states of  $^6$He$^*$, $^3$H$^*$, and $^6$Li$^*$ are formed at the first step  of a two-body reaction, respectively, and then the $^6$He$^*$  excited states (or, analogously, the  $^3$H$^*$ or $^6$Li$^*$ excited states)  decay at the second step of the reaction into the corresponding $\alpha$+n+n (or into p+n+n or p+$\alpha$+n, respectively) three-body channel. In process (\ref{eq1}) the p charged-particle  is the spectator, while in process (\ref{eq2}) the spectator role is played by the $\alpha$ particle.
In addition, as  is well known and also reported in\cite{Furic}, in   bidimensional coincidence spectra (like our observed ($E_{\rm p}$, $E_{\rm \alpha}$) spectra)  of a kinematically incomplete experiment  leading to a four-body reaction, it is possible to observe in ($E_{\rm p}$, $E_{\rm \alpha}$)  bidimensional spectra a correlation between the $E_{\rm p}$ and $E_{\rm \alpha}$ values of registered coincidence events only when the spectator particle comes from the   first step of a two-body reaction mechanism and one particle of the subsequent three-body decay is detected  (see for example processes (\ref{eq1}) and (\ref{eq2})).
In the case of process (\ref{eq3}), since we do not detect neutrons in our experiment,  one neutron is the spectator at  the first step of the two-body reaction while at the second step another neutron is produced, in the decay of the $^6$Li$^*$ excited state into  a three-body p+$\alpha$+n formation. Therefore, p-$\alpha$ coincidence events due to process (\ref{eq3}) cannot give any appreciable correlation between  $E_{\rm p}$ and $E_{\rm \alpha}$ values because this process causes   a statistical relative  energy distribution. For example, in the case presented in Fig. \ref{f1},  where the $p$ and $\alpha$ detector telescopes are placed at $\theta_{\rm p}=28.5^\circ$ and $\theta_{\rm \alpha}=16.5^\circ$, and by considering that the  $E_{\rm \,^4He}$ beam energy is 27.2 MeV, the highest $^6$Li$^*$  excited state  is at $E^*=5.366$~MeV and the  kinetic energy of the formed $^6$Li$^*$ nucleus is 6.9 MeV. The $E_p$ energy in the laboratory system of protons emitted at decay of $^6$Li$^*$ ($E^*=$~5.366 MeV) into $\alpha$+p+n does not exceed 3.7 MeV (see arrow at $E_{\rm p}=3.7$~MeV in Fig. \ref{f1}). 
Consequently,  the contribution of  p-$\alpha$ coincidence events due to process (\ref{eq3}) can affect the events produced by process (\ref{eq1}) in the $3.2-3.7$~MeV $E_{\rm p} $ energy region only. But, such a possible effect would not influence the results of the $E^*$ and $\Gamma$ measurements of $^6$He$^*$, as will be explained in Sec. 3.  

Processes (\ref{eq4}) and (\ref{eq6}) are, in  general,  both three-body formation at the first step of the reaction consisting of  unbound resonance states $^5{\rm Li}^*$ and $^5{\rm He}^*$,  with  n+n and p+n pairs of particles, respectively. In the second step of the reaction the $^5{\rm Li}^*$ and $^5{\rm He}^*$ excited nuclei decay into the p+$\alpha$ and n+$\alpha$ two-body channels, respectively. Therefore,  in the above-mentioned cases,  processes  (\ref{eq4}) and (\ref{eq6})   do not produce any appreciable correlation between the $E_{\rm p}$ and $E_{\rm \alpha}$ values of the  coincidence event yields due to the  statistical relative   energy  distribution of  the remaining two neutrons in the four-body reaction of process (\ref{eq4}), and  proton and neutron particles in the case of process  (\ref{eq6}). In fact, in both of these  processes, the two particles at the  first step of the reaction (the two neutrons, or the proton + neutron) do not have the role of a single spectator. Moreover, since the energy of protons in process   (\ref{eq3}) does not exceed 3.7 MeV, a fortiori   protons produced in process  (\ref{eq4}) cannot exceed such energy value because they come from a three-body reaction at first step. In addition, in the case of process (\ref{eq6}), if the detected proton energy produced at the first step of the three-body reaction overcomes 3.7 MeV the $\alpha$ particle coming from the  $^5{\rm He}^*$ decay can not reach a kinetic energy higher than the 4.4~MeV necessary to be detected by the $\alpha$ telescope, because the energy is carried out by the neutron produced in the first step of the process.
Therefore, also this process does not affect the ($E_{\rm p}$, $E_{\rm \alpha}$) coincidence events. Finally, the process (\ref{eq8}) is the direct statistical four-body break-up and it  produces a flat low background contribution corresponding to a total value of coincidence events lower than about 10\% of the complete set of registered ($E_{\rm p}$, $E_{\rm \alpha}$) coincidences analyzed in each spectrum obtained as a  projection of the p-$\alpha$ coincidence yields onto the $E_{\rm p}$ energy axis.

 The yield of each process depends on the kinematic conditions of reacting nuclei and the geometric configuration of detectors because the phase-space factor is determined by these conditions. To  calculate  the phase-space factor we use the procedure presented in the paper\cite{Furic}. With respect to the study of Furic and Foster\cite{Furic}, the yields of the p-$\alpha$ coincidence events, obtained by analysis of each ($E_{\rm p}$, $E_{\rm \alpha}$) bidimensional  spectrum, were projected onto the $E_{\rm p}$-axis in order to have information about the formed $^6{\rm He}^*$ resonance states. In fact, in the case of the p-$\alpha$ coincidence event detection, if p is the spectator particle (the residual nonresonant particle) and $\alpha$ is the  particle constituting the $^6{\rm He}^*$ three-body resonance, the observation of the population of such a resonant state can be made by projecting the yields of the  ($E_{\rm p}$, $E_{\rm \alpha}$) bidimensional spectrum on the $E_{\rm p}$-axis. In such a case, the energy of the nonresonant particle for a given angle is completely determined by that angle and the excitation energy of the three-particle subsystem. Therefore, a $^6$He$^*$  three-body resonance formation appears as a strip parallel to the $E_{\rm \alpha}$ energy axis. The projection of the p-$\alpha$  coincidence event yields onto the  $E_{\rm p}$-axis produces an energy spectrum that gives information about the formed $^6{\rm He}^*$ excited states decaying into the $\alpha$+n+n three-body channel  by   process (\ref{eq1}). The projection of the coincidence event yields onto the $E_{\rm \alpha}$-axis produces a spectrum that can give information about the  formed $^3{\rm H}^* $ excited states decaying into the p+n+n three-body channel by  process (\ref{eq2}), but in our studied ($E_{\rm p}$, $E_{\rm \alpha}$) bidimensional spectra the contribution of this process cannot be present. In fact, in all of the investigated spectra obtained by the various $\theta_{\rm p}$, $\theta_{\rm \alpha}$ geometric configurations of telescopes,
the very wide  $E_{\rm \alpha}=4.4-14$~MeV energy interval of the spectator $\alpha$-particle of process (\ref{eq2}) corresponds to $E_{\rm (p-2n)}$ relative energies not higher than the 3 MeV of the $^3$H$^*$ excitation energy. Therefore, since the experimental threshold energy is 8.482 MeV (as reported in\cite{attila}) for the p+n+n three-cluster formation in the decay of $^3$H$^*$ excited state, in our investigated bidimensional spectra no contribution of events due to process (\ref{eq2}) can be present.  Moreover, the statistical direct four-body breakup contribution is considered as a flat low background contribution in our analyzed spectra. 

The motion of the two outgoing neutrons can be represented as the motion of their center-of-mass and the relative motion of these two neutrons. The events that correspond to relative energy of two emitted neutrons equal to zero can be considered as events of a p+$\alpha$+$<2n>$ three-body reaction where $<2n>$ represents the dineutron cluster formed by two neutrons which move with the same velocity and direction,  and the kinematic loci  of p-$\alpha$ coincidence events connected with the $^6{\rm He}^*$ and $^3{\rm H}^* $ formations are placed along the solid white lines  (upper or lower branches) reported in the  ($E_{\rm p}$, $E_{\rm \alpha}$) bidimensional spectrum (see Fig \ref{f1}). By increasing the relative energy of the two neutrons, the events are distributed through the allowed kinematic  area of the ($E_{\rm p}$, $E_{\rm \alpha}$)-plane for the four-body reaction, inside the region delimited by the contour of the three-body reaction loci. 

On the basis of the previous kinematic analysis related to the processes (2)--(6), we can affirm that in Fig. \ref{f1}   the experimental p-$\alpha$ coincidence events, represented by red dots, mainly consists of process (\ref{eq1}) where the proton is the spectator and the $\alpha$ is the charged particle coming from the decay of the $^6$He$^*$ excited states into the $\alpha$+n+n three-body cluster. 
The arrow  at $E_{\rm p}=3.2$~MeV in Fig. \ref{f1} indicates the lower limit of proton energy detection by the telescope placed on the right side. Light grey
dots represent the  result of a simulation connected with the upper branch of kinematic loci for the p+$\alpha+<$2n$>$ reaction products when the finite resolution of the detector system and beam energy are considered. Blue dots represent the simulation of process (\ref{eq8}).

Therefore, at the conclusion of the above-mentioned description of the processes (\ref{eq1})--(\ref{eq8}), only process  (\ref{eq1}) of the  $^4$He$+ ^3$H four-body reaction can in principle produce  enhancements in the analyzed spectra obtained as a projection of the p-$\alpha$ coincidence yield vs. $E_{\rm p}$ in connection with the formation of the $^6{\rm He}^*$ and $^3{\rm H}^*$ excited states, respectively.

\section{Analysis and Results}

In the present experiment, by considering that the thickness of 100 $\upmu$m  of the $\Delta E$-detector devoted to the detection of protons does not allow us to measure protons with energy lower than 3.2~MeV, we may study the $E^*$ excitation energy spectrum of $^6$He from the 1.45 to 3.65~MeV energy range in connection with the $\Delta E-E$ telescope placed at $\theta_{\rm p}=28.5^\circ$, and from the 1.1 to 3.1~MeV energy range for the detector placed at $\theta_{\rm p}=36^\circ$. 
In Fig. \ref{f1}, it is evident the cut of experimental events at $E_{\rm \, p}< 3.2$~MeV, while the allowed region populating the p-$\alpha$ coincidence  of the four-body reaction lies within the kinematic loci of the $^3$H($^4$He,p$\alpha$)$<2n>$ three-body reaction when the two neutrons are considered as one particle (dineutron).

\begin{figure}
\centering{\resizebox{1.\textwidth}{!}{\includegraphics{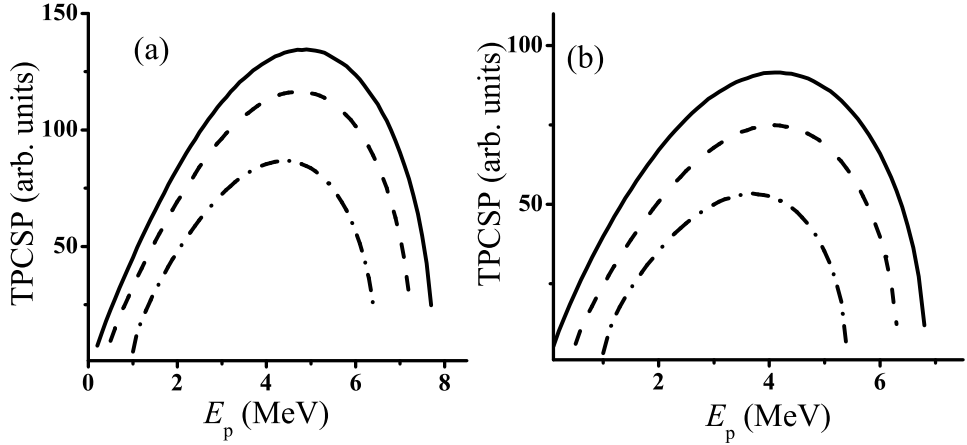}}}
\caption{Two-particle (p-$\alpha$) coincidence phase-space factor $\rho_{_{\rm TPCPS}}$  versus the emitted proton energy $E_{\rm p}$ for the $^3$H($^4$He,p$\alpha$)2n four-body reaction, at the $E_{\,\rm ^4He}$ beam energy of 27.2 MeV and the proton detector placed at the angle $\theta_{\rm p}=28.5^\circ$ in (a) and $\theta_{\rm p}=36^\circ$ in (b). Full, dashed and dash-dotted lines  correspond to $\alpha$-particle detector placed at 13$^\circ$, 16.5$^\circ$, and 19.5$^\circ$, respectively.}\vspace{-0.2cm}
\label{f2ab}       
\end{figure}

Resonances in  two-body and three-body subsystems cause an increase in the intensity of break-up events in those places where the corresponding energy of the relative motion of decayed particles achieves resonance energies. 
In our obtained experimental data we did not observe any resonance phenomenon caused by the p-$\alpha$ interaction (formation of $^5$Li$^*$ by process (\ref{eq4})), while in all registered ($E_{\rm p}$,~$E_{\rm \alpha}$) bidimensional spectra one can observe the strips parallel to the $E_{\alpha}$-axis consisting of events from  process (\ref{eq1}). These strips have been identified as a manifestation of the first $^6$He$^*$ excited state at $E^*=1.797$~MeV and its subsequent decay into the three $\alpha$+n+n components  for which the threshold energy is 0.974 MeV. We assume that besides the narrow first known excited $^6$He level, broader resonance structures at lower proton energies may populate the ($E_{\rm p}$,~$E_{\rm \alpha}$) bidimensional spectra . If we project all p-$\alpha$ coincidence events of the $^3$H($^4$He,p$\alpha$)2n four-body reaction onto the $E_{\rm p}$ axis, connected with the proton detector placed at $\theta_{\rm p}$ and $\alpha$ detector placed at $\theta_{\rm \alpha}$, we obtain the individual experimental spectrum of the p-$\alpha$ coincidence events $N$ versus $E_{\rm p}$,  related to the two-particle coincidence phase-space  (TPCPS): 

\begin{equation}
N \propto \rho_{_{\rm TPCPS}} (\Omega_{\rm p}, \Omega_{\rm \alpha},E_{\rm p})
\end{equation}
where $\rho_{_{\rm TPCPS}}(\Omega_{\rm p}, \Omega_{\rm \alpha},E_{\rm p})$ is the projection onto the $E_{\rm p}$ axis of the phase-space  ratio  for the detection of  two coincidence particles (proton and $\alpha$-particle) from the $^3$H($^4$He,p$\alpha$)2n four-body reaction. 
We calculated the energy dependence of the projection of phase-space  $\rho_{_{\rm TPCPS}}$ for a fixed proton  detection angle and various $\theta_{\rm \alpha}$ angles of the alpha-detector following the papers \cite{Suzuki,Furic} (see Fig. \ref{f2ab}).

Figure 3 (a) and (b) show that the shapes of the $\rho_{_{\rm TPCPS}}(E_{\rm p})$ distributions are similar when $\theta_{\rm p}$ is fixed and $\theta_{\rm \alpha}$ ranges through a set of values, even if the peak energy of each phase-space distribution changes slightly for the lines represented in figures (a) and (b).
Such a behavior allows one to add   several individual experimental energy spectra obtained for a fixed $\theta_{\rm p}$ angle and  various $\theta_{\rm \alpha}$ angles in one  cumulated spectrum, in order to check if in such a cumulated spectrum the  energy distribution  of events $N(E_p)$ keeps the same feature of the individual spectra with respect to the observed low-lying $^6$He$^*$ excited states.   Of course, the final result of the $E^*$ peak energy and $\Gamma$ energy width determinations of the investigated $^6$He$^*$ resonances can be slightly affected by the procedure of cumulated spectra. For this reason we want to check if two close peaks in individual spectra are still well resolved in the cumulative spectrum. If this occurs, such a result confirms that the two close peaks observed in the individual spectra do not arise from fluctuations of data. In fact, also in the presence of coincidence events coming from $\alpha$ particles emitted within an angular range of about 6$^\circ$, the results  found for the two close  $^6$He$^*$ excited states by analysis of individual spectra are confirmed by the results found in the cumulative spectrum obtained for coincidence events registered at the same $\theta_{\rm p}$ angle of spectator proton but at various $\theta_{\alpha}$ angles of $\alpha$ particle emitted in the decay of $^6$He$^*$ in process (\ref{eq1}).

\begin{figure}
\centering{\resizebox{1.\textwidth}{!}{\includegraphics{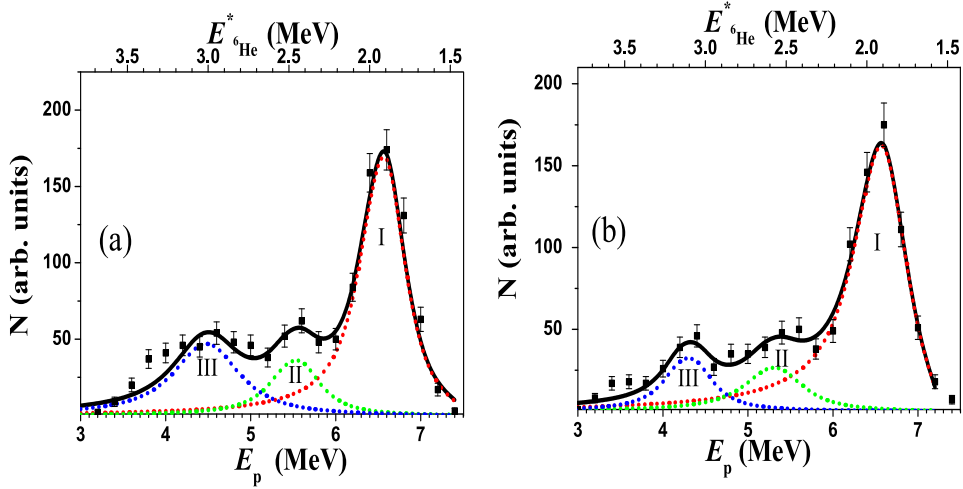}}}
\caption{(Color online) The individual spectrum obtained as a projection of the p-$\alpha$ coincidence yields onto the $E_{\rm p}$ energy axis connected with the detector placed at $\theta_{\rm p}=28.5^\circ$  registered for the $^3$H($^4$He,p$\alpha$)2n four-body reaction for the $E_{\rm ^4He}$ beam energy of 27.2~MeV. Dotted lines represent the contributions of the various $^6$He$^*$ resonance states   with the labels I, II, and III, while the full line represents the sum of all resonance contributions. In the upper scale  the $E^*_{\rm ^6He}$ excitation energy values are reported. (a) The spectrum obtained for the $\alpha$-particle detector placed at  $\theta_{\rm \alpha}=13^\circ$; (b) the spectrum for $\theta_{\rm \alpha}=16.5^\circ$. }
\label{f3ab}       
\end{figure}

In the calculation we take into account the beam energy and its dispersion, the  target thickness, the energy loss in the target and the energy resolution of the  detectors.
Figure  \ref{f3ab} shows the individual energy spectrum $N(E_{\rm p})$ vs. $E_{\rm p}$ of the p-$\alpha$ coincidence events projected onto the $E_{\rm p}$ axis,  obtained for telescopes placed at  $\theta_{\rm p}=28.5^\circ$, and  $\theta_{\rm \alpha}=13^\circ$ (see Fig. \ref{f3ab} (a)) and $\theta_{\rm \alpha}=16.5^\circ$ (see Fig. \ref{f3ab} (b)).
In each figure  three peaks clearly appear and we can fit the data with the sum of the following Breit-Wigner contributions

  \begin{equation}
   N \propto \rho_{_{TPCPS}}(\Omega_{p}, \Omega_{\alpha}, E_{\rm p})   \times  \sum^3_{j=1} C_j \frac{(1/2\Gamma_j)^2}{{(E_{j}-E_{\,(\alpha-2n)})}^2 + {(1/2\Gamma_j)}^2} 
   \label{eq9}
  \end{equation}
where $\rho_{_{\rm TPCPS}}$ is the phase-space factor calculated with the Monte Carlo simulation of the p and $\alpha$ detected particles from the $^3$H($^4$He,p$\alpha$)2n four-body reaction, $C_{\rm j}$ is the yield of the corresponding contribution of each  $^6$He$^*$ resonance state decaying into $\alpha$+n+n particles, and $E_{\,(\alpha-2n)}$ is the ($\alpha$-2n) relative energy value at decay of $^6$He$^*$ excited state into $\alpha+$2n products. This relative energy is uniquely determined by the  $\theta_{\rm p}$ angle and $E_{\rm p}$  proton energy  in the case in where the proton is the nonresonant particle leaving the $^6$He$^*$ excited nucleus at the first step of the reaction,  while the $\alpha$-particle is the resonant particle obtained at decay of $^6$He$^*$ into the $\alpha$+n+n particles. Moreover, $E_j$ is the peak energy value of each $^6$He$^*$ resonance state and $\Gamma_j$ is its energy width.

The solid line represents the sum of the three $^6$He$^*$ resonance state contributions described by the dashed lines. The results of the $E^*$ and $\Gamma$ for the three $^6$He$^*$ resonance states present in Figs. \ref{f3ab} (a) and (b) are given in Table \ref{tb1}.
It must be  considered that the resolution of $E^*$ peak energy and $\Gamma$ energy width in the present work is 0.2~MeV, which is limited by the energy spread of beam and the resolution of the detection system.  For this reason, by investigation of our analyzed bidimensional coincidence spectra of a kinematicaly incomplete (four-body) reaction, we find for the narrow first excited  $^6$He$^*$ state a $\Gamma$ width of at least 0.2 MeV, but the energy resolution of our experimental apparatus allows us to resolve the two close  $^6$He$^*$ excited states at about $E^*=2.4$ and $2.9$~MeV, with $\Gamma=0.3$ and $0.4$ MeV, respectively. 
Therefore, the measure of the well known $^6$He$^*$ excited  state at  1.8 MeV   is a test of the reliability of our analysis.

\begin{table}[h]
\tbl{$E^*$ excitation energy and $\Gamma$ energy width values of the $^6$He$^*$  levels populated by the    $^3$H($^4$He,p$\alpha$)2n  reaction for different geometric detector configurations, using  the Breit-Wigner approximation in (\ref{eq9}).}
{\begin{tabular}{@{}ccccc@{}} \toprule
 $\theta_{\rm p}$ , $ \theta_{\rm \alpha}$  & peak label &$E^*$ (MeV) & $\Gamma$ (MeV) &   see Fig.  \\
\colrule
 28.5$^\circ$  , 13.0$^\circ$  &      I   & 1.9 $\pm$  0.2 &  0.2 $\pm$  0.2 &  \ref{f3ab}~(a) \\
                                  &     II   & 2.5 $\pm$  0.2 &  0.3 $\pm$  0.2 &  '' \\
                                  &    III   & 3.0 $\pm$  0.2 &  0.4 $\pm$  0.2 &  '' \\\hline

 28.5$^\circ$ , 16.5$^\circ$  &     I  & 1.9 $\pm$  0.2 &  0.3 $\pm$  0.2 & \ref{f3ab} (b) \\
                                  &   II & 2.6 $\pm$  0.2 &  0.4 $\pm$  0.2 & '' \\
                                  &     III& 3.1 $\pm$  0.2 &  0.3 $\pm$  0.2 & '' \\ \hline
 36.0$^\circ$ , 19.5$^\circ$ &     I  & 1.8 $\pm$  0.2 &  0.3 $\pm$  0.2 & \ref{f4}  \\
                                 &     II & 2.3 $\pm$  0.2 &  0.3 $\pm$  0.2 & '' \\
                                 &      III& 2.7 $\pm$  0.2 &  0.4 $\pm$  0.2 & '' \\ \botrule
\end{tabular}\label{tb1} }
\end{table}

\begin{figure}[h]
\vspace{0cm}
\centering{\resizebox{0.55\textwidth}{!}{\includegraphics{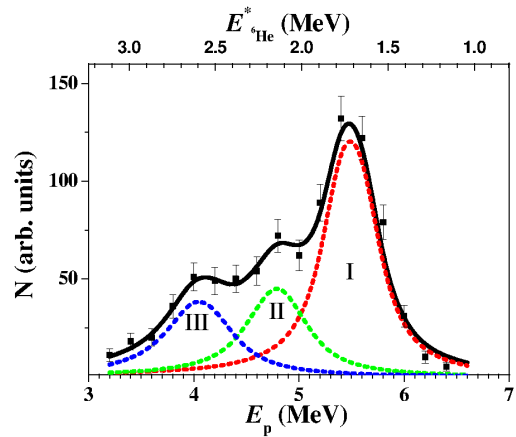}} \vspace{0cm}}
\caption{(Color online) As Fig. \ref{f3ab} (a), but for $\theta_{\rm p}=36^\circ$ and $\theta_{\rm \alpha}=19.5^\circ$. }
\label{f4}       
\end{figure}

Figure \ref{f4}  shows the individual energy spectrum obtained for $\theta_{\rm p}=36^\circ$ and $\theta_{\rm \alpha}=19.5^\circ$. Also in this figure three peaks are   clearly present and we performed calculation by the same procedure used for the data analysis of Figs. \ref{f3ab} (a) and  (b). The solid line represents the sum of the three $^6$He$^*$ resonance contributions. The obtained $E^*$ energy peak and $\Gamma$ energy width values are  reported in Table \ref{tb1}. As one can see, the results listed in Table \ref{tb1} are consistent within the estimated  measurement uncertainties for the analyzed individual energy spectra. In all ($E_{\rm p}$, $E_{\rm \alpha}$) analyzed spectra, the registered p-$\alpha$ coincidence events have been normalized to the number of events collected by a monitor.

By the analysis of event distribution obtained as projection of coincidence event yields onto the $E_{\rm \alpha}$-axis, we verified that even in these experimental conditions  of telescopes at $\theta_{\rm p}=36^\circ$ and $\theta_{\rm p}=19.5^\circ$ no contribution  of coincidence events due to the $^3$H$^*$ excited state 
formation  by  process (\ref{eq2}) was present in the considered ($E_{\rm p}$, $E_{\rm \alpha}$) bidimensional spectra. 

Finally, as one can see in Figs. \ref{f3ab} and \ref{f4}, the possible contribution of process (\ref{eq3}) in the $E_{\rm p}$ energy region from 3.2 to 3.7 MeV does not  affect the results of $E^*$ and $\Gamma$ obtained for the studied $^6$He$^*$ levels, because such events lie in the tail of the N($E_{\rm p}$) distribution.

\begin{figure}[h]
\vspace{0cm}
\centering{\resizebox{0.55\textwidth}{!}{\includegraphics{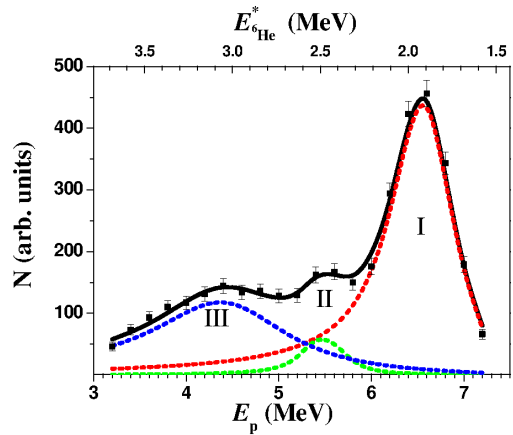}}}
\vspace{-0.1cm}
\caption{(Color online) The cumulative spectrum versus the  $E_{\rm p}$ proton energy  as a sum of the three individual spectra obtained for fixed $\theta_{\rm p}=28.5^\circ$ and the various $\theta_{\rm \alpha}=13.0^\circ$, $\theta_{\rm \alpha}=16.5^\circ$, and $\theta_{\rm \alpha}=19.5^\circ$ angle values. In the upper scale  the $E^*_{\rm ^6He}$ excitation energy values are reported. }
\label{f5ab}       
\end{figure}

In Fig. \ref{f5ab}  we present the cumulative energy spectrum obtained as a sum of the three individual energy spectra collected  at $\theta_{\rm \alpha}=13^\circ\, ,16.5^\circ\mbox{, and } 19.5^\circ$, respectively, when the proton detector is placed at $\theta_{\rm p}=28.5^\circ$.

\begin{table}[h]
\tbl{As Table \ref{tb1}, but for the cumulative spectrum presented in Fig. \ref{f5ab} obtained as a sum of three individual spectra collected at  $\theta_{\rm p}=28.5^\circ$, and  $\theta_{\rm \alpha}=13^\circ\, ,16.5^\circ\mbox{, and } 19.5^\circ$.}
{\begin{tabular}{@{}ccc@{}} \toprule
peak labe & $E^*$ (MeV) & $\Gamma$ (MeV)  \\
\colrule
    I  \,\,& 1.9 $\pm$  0.2 &\,\,\, 0.3 $\pm$  0.1  \\
  II  \,\,& 2.5 $\pm$  0.2 &\,\,\, 0.4 $\pm$  0.2  \\
 III  \,\,& 3.0 $\pm$  0.2 &\,\,\, 0.6 $\pm$  0.2  \\ \botrule
\end{tabular}\label{tb2} }
\end{table}

As one can see,  the $E^*$ and $\Gamma$ values  reported in Table \ref{tb2} for the cumulative energy spectrum are consistent with the respective values presented in Table \ref{tb1} for the individual energy spectra, but only the $\Gamma$ width values are slightly larger due to the  summing of spectra. Even if we subtract in the spectrum of  Fig. \ref{f5ab}  a  flat background  of about 40 events, we do not obtain appreciable  changes of the $E^*$ results reported in Table \ref{tb2}.  In this cumulative spectrum the $^6$He resonant contributions also appear well resolved but wider than the ones observed in the studied   individual spectra. This result demonstrates the small influence on the resonant contributions of coincidence events registered by  detectors placed at $\theta_{\rm p}$ and various  $\theta_{\rm \alpha}$ angles. Then, we can affirm that  the analyzed cumulative spectrum keeps the same feature of the individual spectra. For this reason, we can affirm that the best way to extract the $E^*$ energy peak and $\Gamma$ energy width of low-lying $^6$He$^*$ excited states is  to analyze the individual ($E_{\rm p}$, $E_{\rm \alpha}$) coincidence energy spectra (also much better than analyzing the cumulative spectra) at $\theta_{\rm p}$ and   $\theta_{\rm \alpha}$ detector angles where the  phase-space factor of the reaction process forming the $^6$He$^*$ nucleus in the excited energy range of interest is large. Therefore, we can conclude that in the  investigated  $^3$H($^4$He,p$\alpha$)2n experiment we find and confirm the $^6$He$^*$ first excited state given in literature\cite{selov88,till02} at $E^*=1.8$~MeV decaying into the $\alpha$+n+n particles, but we also find two new close and resolved $^6$He$^*$  excited states decaying into the same $\alpha$+n+n channel.

\begin{figure}[h]
\vspace{-0.5cm}
\centering{\resizebox{0.55\textwidth}{!}{\includegraphics{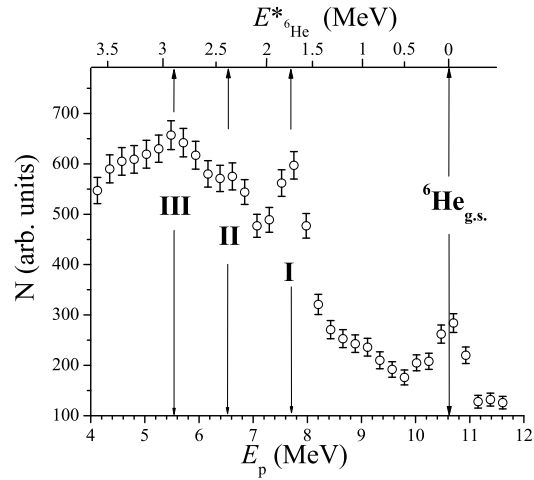}}}
\vspace{-0.1cm}
\caption{Open circles represent the experimental inclusive proton energy spectrum at $\theta_{\rm p}=17^\circ$ of the  $^3$H($^4$He,p)$\alpha$nn  reaction at a beam energy of 27.2 MeV. In the upper  scale   the $^6$He$^*$ excitation energy obtained  by   process (\ref{eq1}) is reported. 
Label I indicates the enhancement corresponding to the first excited state, while labels II and III  indicate the presence of the contributions due to the two new excited states.  
The upper arrows indicate the excitation energies of the $^6$He states, while the lower arrows indicate the corresponding $E_{\rm p}$ proton energies.}
\label{inclusive} 
\end{figure}

In addition to the  individual and cumulative p-$\alpha$ coincidence spectra,  we also show  the inclusive proton energy spectrum of the $^3$H($^4$He,p)$\alpha$nn  reaction  where the detected protons can come from channels (1) and (3)-(6), as explained in Sec. 2.  
Figure \ref{inclusive} shows the  inclusive proton  spectrum at $\theta_{\rm p}=17^\circ$. 
As one can see, such an  experimental $N$($E_{\rm p}$) spectrum clearly exhibits peaks of the $^6$He formation corresponding to: the ground state  (g.s.) with proton kinetic energy of about $E_{\rm p}$= 10.6 MeV, the first excited state with about $E_{\rm p}$= 7.7 MeV, a broad distribution peaked at about $E_{\rm p}$= 5.5 MeV, and an appreciable shoulder at about $E_{\rm p}$= 6.6 MeV.  To these  values of proton kinetic energy correspond on the upper $E_{\rm \,^6He}$ excitation energy axis  the values of about 0 (g.s.), 1.75 (the known 2$^+$ first excited state), 2.9, and 2.4 MeV, respectively. 
 By analyzing this inclusive energy spectrum contributed by all considered allowed channels, it is impossible to resolve close excited states of the $^6$He$^*$ formation and extract reliable  $E^*$ and  $\Gamma$ values. 
From a simple look of Fig. \ref{inclusive}, the $E^*$ values  of $^6$He$^*$ levels   are consistent with the ones obtained in our analyzed   p-$\alpha$ coincidence energy spectra (see Figs. \ref{f3ab} and \ref{f4}). But, due to a  large uncertainty in the background contribution by processes     (\ref{eq1}), and (\ref{eq3})--(\ref{eq8}),  it is not suitable to extract from inclusive spectra reliable  $\Gamma$ width values of the  $^6$He$^*$ levels.

\section{Conclusions}

We performed the $^4$He+$^3$H experiment at $E_{\rm\, ^4He}=27.2$~MeV by using a target with a titanium backing saturated with tritium, with the aim of investigating  the low-lying  $^6$He$^*$  levels populating the energy region just above the threshold energy of the $\alpha$+n+n three-cluster formation.
 By the analysis of the ($E_{\rm p}$, $E_{\rm \alpha}$) energy spectra of the registered p-$\alpha$ coincidence events we find two new well resolved $^6$He$^*$  excited states decaying into the $\alpha$+2n three-cluster channel. From the obtained values for the $E^*$ and the $\Gamma$ energy width  presented in Table \ref{tb1}  for various angle configurations of detectors,  we suggest the following averaged values: $E^*=2.4 \pm 0.2$~MeV with $\Gamma=0.3\pm0.2$~MeV, and $E^*=2.9 \pm 0.2$~MeV with $\Gamma=0.4\pm0.2$~MeV for the two new $^6$He$^*$  resonance states, respectively. 
It is useful to compare the results obtained in the present paper by analyzing  the ($E_{\rm p}$, $E_{\rm \alpha}$) bidimensional spectra of the studied $^3$H($^4$He,p$\alpha$)2n experiment for the excited levels of the $^6$He$^*$ formation in the $E^*$ energy region from 1.45 to 3.65 MeV with the results found by   Mougeot {\it et al.} \cite{mougeot12} in the same $^6$He$^*$ energy region by the analyzed triton-particle spectra of the investigated p($^8$He,t) reaction at $^8$He beam energy of 15.4~MeV/nucleon. By the fit of the $N$($E_{\rm p}$) distribution we extract the $E^*$ and $\Gamma$ values of the first excited state and two new 2.4 and 2.9 MeV  levels of $^6$He$^*$  with $\Gamma=$0.3 and 0.4 MeV, respectively. Instead, the authors \cite{mougeot12} in their fit of experimental data used the $E^*$ and $\Gamma$ values given in literature\cite{selov88,till02} for the first $^6$He$^*$ excited state, and they found only one $^6$He$^*$ resonant state at $E^*=$2.6 MeV with $\Gamma=$1.6 MeV, by analyzing the triton-particle spectra with $\theta_{\rm c.m.}$ angle intervals of at least $10^\circ - 20^\circ$ (or even more).
 We think that for the nature of the investigated reaction and the sensitivity of analysis the authors \cite{mougeot12}  observed a wider convolution of the two $^6$He excited states than we found. The appropriate experiment at lower $^4$He beam energy and detailed analysis of the individual ($E_{\rm p}$, $E_{\rm \alpha}$) energy spectra  which we considered are necessary in order to find and well resolve the two close 2.4 and 2.9 MeV $^6$He$^*$  excited states.
 
 The reliability of   spectral analysis and the results found in the present experiment for the considered four-body reaction is confirmed by the values of $E^*=1.8 \pm 0.2$~MeV with $\Gamma=0.3\pm0.2$~MeV obtained for the first $^6$He$^*$ excited state which are consistent   with the values  given in literature \cite{selov88,till02} within the experimental resolution. 
In addition, we also find consistent $^6$He$^*$ level energies by looking the inclusive proton energy spectrum that confirms the presence of the two new low-lying  $^6$He$^*$  excited states decaying into the $\alpha$+2n channel, but we note that it is not suitable to measure the $E^*$ and $\Gamma$ spectroscopic parameters by the inclusive proton spectra.
 Moreover, the two new and close low-lying levels found for the $^6$He$^*$ resonance states decaying into the $\alpha$+n+n three-cluster channel will be useful as a test for the choice of the nucleon-nucleon potential in the theoretical model. In addition, the detailed knowledge of the low-lying excited states of neutron rich light nuclei, such as the $^6$He nucleus, constitutes a basic test to constrain the theoretical models.
 On the other hand, it is also interesting to promote experimental investigations of the low-lying excited states of $^6$Be.

\section*{Acknowledgments}

The authors wish to thank the staff of the Institute for Nuclear Research Laboratories (Kiev) for their help during the measurements. This work was supported by INR of the National Academy of Sciences of Ukraine, and partially by the Istituto Nazionale di Fisica Nucleare in Italy.
We wish to thank Dr. Anthony Palladino, of the INFN-LNF of Frascati, for the English improvement of the text.

\end{document}